\documentclass[
aps,reprint,showpacs,
amsmath,amssymb]{revtex4-1}
\usepackage{graphicx}
\usepackage[colorlinks]{hyperref}
\newcommand*\ee{e}
\newcommand*\ii{i}
\newcommand*\sgn{\operatorname{sgn}}
\newcommand*\Pf{\operatorname{Pf}}
\newcommand*\ket[1]{\mathinner{|{#1}\rangle}}
\newcommand*\PT{\ensuremath{\mathcal{PT}}}

\begin{document}

\title{Topological quantum wires with balanced gain and loss}
\author{Henri Menke}
\email{h.menke@fkf.mpg.de}
\author{Moritz M. Hirschmann}
\email{m.hirschmann@fkf.mpg.de}
\affiliation{Max-Planck-Institut f\"ur Festk\"orperforschung,
  Heisenbergstra\ss e 1, D-70569 Stuttgart, Germany}

\begin{abstract}
  We study a one-dimensional topological superconductor, the Kitaev
  chain, under the influence of a non-Hermitian but \PT-symmetric
  potential.  This potential introduces gain and loss in the system in
  equal parts.  We show that the stability of the topological phase is
  influenced by the gain/loss strength and explicitly derive the bulk
  topological invariant in a bipartite lattice as well as compute the
  corresponding phase diagram using analytical and numerical methods.
  Furthermore, we find that the edge state is exponentially localized
  near the ends of the wire despite the presence of gain and loss of
  probability amplitude in that region.
\end{abstract}


\maketitle

\section{Introduction}

Since the successful fabrication of topological insulators
\cite{Koenig2007} and superconductors \cite{Nadj-Perge2014} in the
last decade, enormous progress has been made in understanding and
optimizing these non-trivial topological phases of matter.  The first
reported topologically non-trivial state, the integer quantum Hall
effect \cite{Klitzing1980}, exhibits a fully gapped bulk with gapless
chiral edge states.  It is distinct from other states like the quantum
spin Hall effect in the sense that it does not rely on any symmetries
of the Hamiltonian.  The quantum spin Hall state is only robust
against perturbations which do not break time-reversal symmetry
\cite{Bernevig2006}.  This symmetry-protection can be generalized and
has been summarized in periodic tables of topological insulators and
superconductors \cite{Schnyder2008, Kitaev2009, Ryu2010, Chiu2016,
  Zhao2016}.  Topological superconductors are particle-hole-symmetric
and exhibit gapless surface states.  The particle- and hole-like
excitations are an analogy to particle and anti-particle pairs which
allows a description in terms of Majorana fermions \cite{Kitaev2001}.

Symmetries of the Hamiltonian operator are relevant in the field of
non-Hermitian quantum mechanics \cite{Moiseyev2011, Konotop2016}.  The
postulates of quantum mechanics demand that observables are
represented by Hermitian operators which have real eigenvalues.  It
was shown by \citet{Bender1998} that the weaker constraint of
parity-time (\PT) symmetry is sufficient for an operator to have a
purely real eigenspectrum.  Still, the eigenstates of an \emph{a
  priori} \PT-symmetric Hamiltonian can spontaneously break the
\PT~symmetry \cite{Bender2007}.  These non-Hermitian systems play an
important role in physics and have been studied in the context of
localization-delocalization transitions of flux lines in type-II
superconductors \cite{Hatano1997}, the disordered Anderson model
\cite{Goldsheid1998}, dissipative quantum systems \cite{Rudner2009,
  Diehl2011}, and most recently in topological insulators
\cite{Hu2011, Kenta2011, Zhu2014, Lee2016, Leykam2017} and topological
superconductors \cite{Wang2015, Yuce2016, Klett2017, SanJose2016}.
Experimental realizations have been achieved in photonic lattices and
photonic crystals \cite{Makris2008, Guo2009, Rueter2010, Bittner2012}.

An open quantum system where the probability amplitude is not
conserved and which is subject to in and out flux in the time
evolution but still supports stationary solutions is called a system
with balanced gain and loss.  Gain and loss effects are usually
studied with the non-Hamiltonian approach of a Lindblad master
equation \cite{Breuer2002}.  In a mean-field description this can be
substituted by imaginary potentials fulfilling \PT~symmetry, as has
been studied for Bose-Einstein condensates \cite{Trimborn2008,
  Dast2016}.

In this paper, we consider an extension of the well-known Kitaev chain
\cite{Kitaev2001} using \PT-symmetric potentials to introduce balanced
gain and loss effects.  This model has been investigated before in the
context of the interplay between \PT~symmetry breaking and the
topological phase \cite{Wang2015, Yuce2016, Klett2017}.  We
analytically derive the topological invariant given by the Pfaffian
for a specific choice of the potential and compare it to numerical
results.  Furthermore, we study the localization properties of the
edge state in the topological regime by the generating function
approach \cite{Patrick2016}.  Using this method, we analytically
compute the decay constant of the edge state in the bulk and find a
criterion for the topological phase transition.

\section{Kitaev Chain with Gain and Loss}

The symmetries $\mathcal{P}$ and $\mathcal{T}$ are defined as the
space-reflection (parity) and time-reversal operator with the actions
$t\to t$, $x\to-x$, $\ii\to\ii$ and $t\to-t$, $x\to x$, $\ii\to-\ii$,
respectively with $t$ denoting time and $x$ denoting position.  In the
discrete lattice case these actions can be described by $\mathcal{P}
c_n \mathcal{P} = c_{N+1-n}$ and $\mathcal{T} \ii \mathcal{T} = -\ii$
with annihilation (creation) operators $c_n$ ($c_n^\dagger$).  A
Hamiltonian operator is considered to be \PT-symmetric if it commutes
with the union of the $\mathcal{P}$ and $\mathcal{T}$ operator
$[\PT,H]=0$.  It is not necessary that $H$ commutes with either of the
operators alone.

Let us consider an in general non-Hermitian system which has
\PT~symmetry.  Furthermore it is subject to particle-hole symmetry
which means the Hamiltonian fulfills $H_{\text{PHS}} = - \tau_1
H^T_{\text{PHS}} \tau_1$ with $\tau_1$ denoting the first Pauli
matrix.  \PT~symmetry of a Hamiltonian $H$ implies $O^\dagger H^* O =
H$ with a unitary matrix $O$.

The presence of particle-hole symmetry allows us to apply a basis
transformation $M$ to Majorana operators which relates the matrix $H$
via $H = (\ii/4) M^\dagger X M$ to $X$ which is skew-symmetric.  Under
the change of basis the commuting \PT~symmetry $[\PT,H] = 0$ becomes
an anti-commuting one $\{U,X\} = 0$ , where $U$ is an antiunitary
symmetry (see Appendix~\ref{app:invariant}).  For the simplest case of
an orthogonal matrix $O$ and vanishing diagonal elements in $X$ we
obtain
\begin{equation}
  \Pf(X)^* = \Pf(O X O^T) = \Pf(X) \det(O) = \Pf(X) \;,
\end{equation}
i.e., a real Pfaffian.  For a more detailed and general discussion we
refer to the Appendixes.  Therefore, we have shown that the sign of
the Pfaffian is well defined in certain non-Hermitian systems with
\PT~symmetry. The sign of the Pfaffian is known to be related to the
$\mathbb{Z}_2$ topological invariant for the symmetry class D of
Hermitian systems.  Accordingly, we suggest that the same invariant
may also be used to classify these \PT-symmetric non-Hermitian
systems.  We will elaborate this statement with an example where the
continuity between a Hermitian and a non-Hermitian system is evident.

In the following, we consider a system described by a non-Hermitian
Hamiltonian $H$ with \PT~symmetry.  It consists of a Hermitian part
$H_0$ which is in our case given by the Kitaev chain Hamiltonian and a
non-Hermitian $U$ which commutes with the \PT~operator.  In general,
one has
\begin{equation}
  H = H_0 + U \;,\quad U \neq U^\dagger \;.
\end{equation}
We will refer to this full Hamiltonian as the extended Kitaev chain in
this paper.  The Kitaev chain Hamiltonian was introduced in
\cite{Kitaev2001} as a pedagogical model to describe topological
superconductivity in a chain of spinless fermions.  The Hamiltonian
reads as
\begin{equation}
  \label{eq:kitaev}
  H_0 = \sum_n \left[ t c_n^\dagger c_{n+1} + \Delta c_n c_{n+1}
    - \frac{\mu}{2} c_n^\dagger c_n + \mathrm{H.c.} \right]
\end{equation}
with the hopping amplitude $t$, the $p$-wave pairing parameter $\Delta
= |\Delta| \ee^{\ii\theta}$, and the chemical potential $\mu$.  The
operators $c_n$ ($c_n^\dagger$) are the fermionic annihilation
(creation) operators of quasiparticles.  In general, the Hamiltonian
\eqref{eq:kitaev} is not \PT-symmetric, but for a choice of the
superconducting phase $\theta = \pm\pi/2$ it is \cite{Wang2015}.
Therefore, we write \eqref{eq:kitaev} as
\begin{equation}
  \label{eq:kitaev-real}
  H_0 = \sum_n \left[ t c_n^\dagger c_{n+1} + \ii \Delta c_n c_{n+1}
    - \frac{\mu}{2} c_n^\dagger c_n + \mathrm{H.c.} \right] \;,
\end{equation}
where the gap parameter $\Delta$ is real.

We investigate two choices for the potential $U$.  One simple
\PT-symmetric potential is given by alternating gain and loss at each
site.  This potential is chosen because the full Hamiltonian can then
be diagonalized analytically by enlarging the unit cell.  The
expression reads
\begin{equation}
  \label{eq:bipartite}
  U_1 = \sum_n (-1)^n \ii \gamma c_n^\dagger c_n \;,
\end{equation}
where $\gamma$ is a real number.  The other choice breaks Hermiticity
only in some regions of the chain.  To consider a general way of in-
and out-coupling fermions, we take the complex potential over a
partitioning of $N/2-f$ with $f \in \mathbb{N}_0$ starting from the
edges
\begin{equation}
  \label{eq:partitioning}
  U_2 = \ii \gamma \biggl[ -\sum_{n=1}^{N/2-f} c^\dagger_n c_n
  + \sum_{n=N/2+f}^{N} c^\dagger_n c_n \biggr] \;.
\end{equation}

\section{Topological Invariant}

In previous works \cite{Wang2015,Yuce2016} the influence of the
\PT-symmetry breaking transition on the topological phases of the
Kitaev chain has been studied.  However, the explicit computation of
the topological invariant can be achieved for non-Hermitian systems as
shown, for example, in \cite{Liang2013}.  \citet{Hu2011} state a no-go
theorem for topological insulator phases in \PT-symmetric systems
showing that they cannot have a real eigenvalue spectrum but leave
topological superconductors open for discussion.  Purely real spectra
of topological superconductors in the presence of a non-Hermitian
potential have been discussed in \cite{Wang2015}.

Depending on the gap parameter $\Delta$, the Kitaev chain can belong
to two different Altland-Zirnbauer symmetry classes
\cite{Schnyder2008, Kitaev2009, Ryu2010, Chiu2016}.  For real
$\Delta\in\mathbb{R}$ it belongs to class BDI whereas for complex
$\Delta\in\mathbb{C}$ it is in class D.  For class BDI the invariant
is a winding number and has $\mathbb{Z}$ character.  In class D the
invariant is $\mathbb{Z}_2$ and is determined by the Pfaffian
\cite{Kitaev2001}.

The Pfaffian is calculated by bringing the Hamiltonian itself into a
skew-symmetric form and the topological invariant is determined by the
sign of the Pfaffian.  Now that the Hamiltonian contains non-Hermitian
terms the Pfaffian is not necessarily real.  It can be shown, however
(see Appendix~\ref{app:invariant}), that in case of a non-Hermitian
\PT-symmetric on-site potential the Pfaffian remains a real quantity
which leaves the sign well-defined and renders it eligible as a
topological invariant for the model at hand.

\begin{figure}
  \centering
  \includegraphics{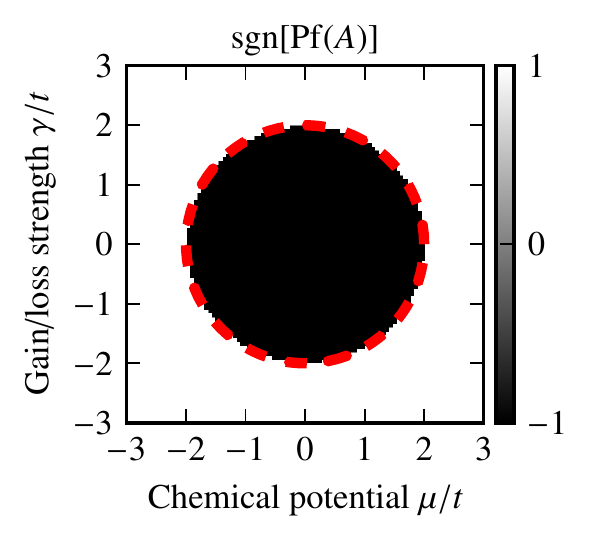}
  \caption{Topological phase diagram for the Kitaev chain with $N =
    100$ sites and the alternating potential \eqref{eq:bipartite} for
    $t=1$ and $\Delta=1$.  The colormap indicates the sign of the
    Pfaffian where $-1$ (black) means topological and $+1$ (white)
    means trivial.  The dashed line corresponds to the analytically
    calculated phase boundary.}
  \label{fig:pd-bipartite}
\end{figure}

For the potential in Eq.~\eqref{eq:bipartite}, the Pfaffian can be
calculated analytically (see Appendix~\ref{app:pfaffian}).  In this
case, the Pfaffian is purely real and the sign of the Pfaffian is well
defined.  In addition, we find a circle criterium for the phase where
the Pfaffian is negative
\begin{equation}
  \label{eq:circle-criterium}
  \mu^2 + \gamma^2 < 4t^2 \;.
\end{equation}
Due to the presence of edge states, we will refer to this as the
topologically non-trivial phase.  The sign of the Pfaffian can also be
extracted from a real-space tight-binding calculation and it is found
that the analytical and the numerical results coincide perfectly.  The
corresponding topological phase diagram is shown in
Fig.~\ref{fig:pd-bipartite}.

\begin{figure}
  \centering
  \includegraphics{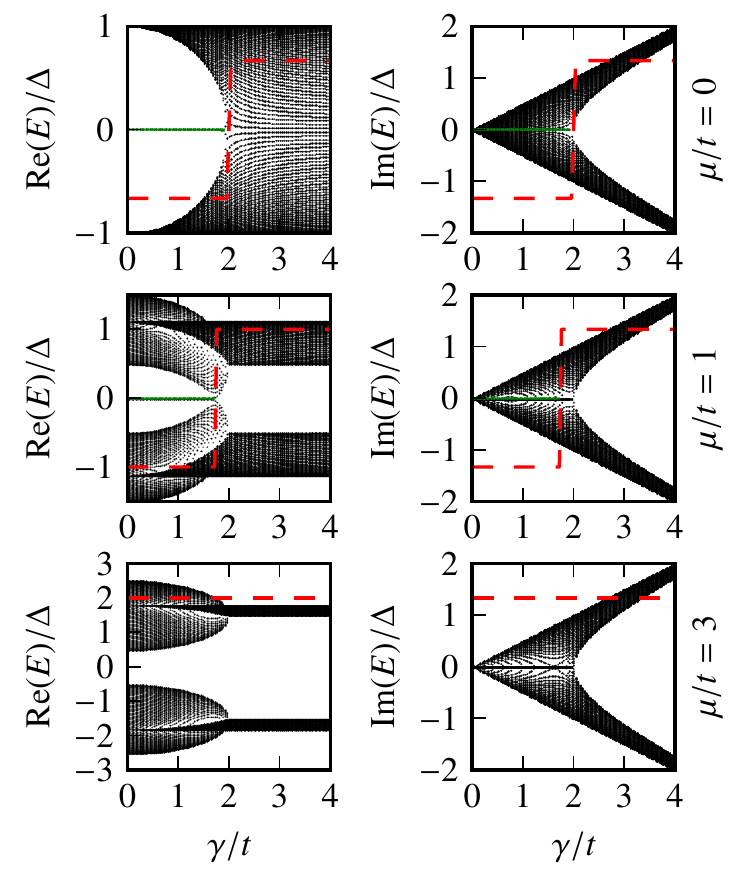}
  \caption{We show spectra for a chain with $N=100$ sites and the
    alternating potential \eqref{eq:bipartite} at different chemical
    potential over the gain/loss strength $\gamma$.  The dashed line
    indicates the sign of the Pfaffian invariant.  Zero modes
    [$\operatorname{Re}(E) = \operatorname{Im}(E) = 0$] are
    highlighted with green dots.  Evidently, the potential can close
    the gap and lead to a topological phase transition.  In the
    topological regime, we always have a state at zero energy.}
  \label{fig:spectra-bipartite}
\end{figure}

\begin{figure}
  \centering
  \includegraphics{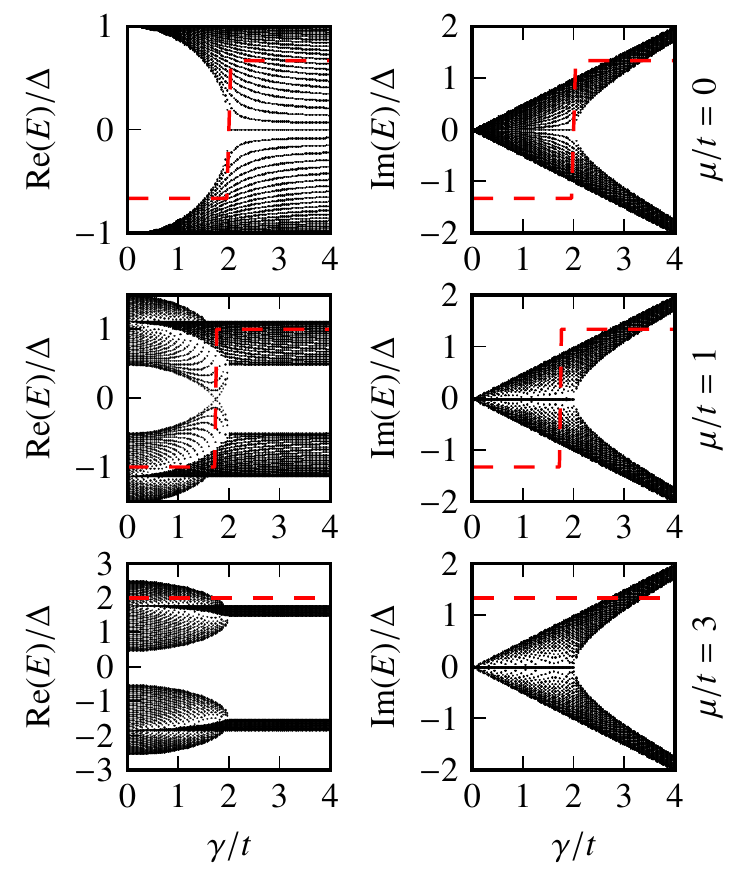}
  \caption{We show bulk spectra for a chain with $N=100$ sites and the
    alternating potential \eqref{eq:bipartite} at different chemical
    potential over the gain/loss strength $\gamma$.  The dashed line
    indicates the sign of the Pfaffian invariant.  For periodic
    boundary conditions the spectra differ qualitatively from open
    boundary conditions.}
  \label{fig:spectra-bipartite-pbc}
\end{figure}

For non-interacting Hermitian systems, the topological phase
transition is accompanied by a gap closing.  Therefore, we compute the
spectrum of the extended Kitaev chain for various choices of the
chemical potential and observe the real and imaginary parts of the
eigenvalues as a function of the gain/loss strength $\gamma$.
Relevant eigenvalue spectra are shown in
Fig.~\ref{fig:spectra-bipartite}.  The \PT-unbroken states of the
system are those with imaginary part $\operatorname{Im}(E)=0$ which do
not decay in the time evolution.  The boundary states with zero energy
are hence \PT-unbroken and robust against gain and loss even during
time evolution.  The bulk spectrum for this system, which we show in
Fig.~\ref{fig:spectra-bipartite-pbc}, differs qualitatively from the
open chain, as has been observed before for a different system in
\cite{Lee2016}.  Parameters $\mu/t = 0$ and $1$ correspond to systems
which can be topologically nontrivial according to
Eq.~\eqref{eq:circle-criterium} while $\mu/t = 3$ always corresponds
to the trivial case.  Figures~\ref{fig:spectra-bipartite}
and~\ref{fig:spectra-bipartite-pbc} show that the gapless state of the
open system is not present in the bulk system as it is expected for
edge states.  We find that at the topological phase transition, the
real part undergoes a bulk gap closing whereas the imaginary part,
which is zero in the topological regime, might open a gap, here for
$\mu = 0$.  The phase transition is characterized by a gap closing in
the modulus of the energy eigenvalue.

\section{Edge State Localization}

Since a complex potential can be interpreted as particles added or
removed from the system, it seems natural to expect a modified spatial
dependence of the edge states.  To verify this expectation we can
apply the generating function method which relates the exponential
decay of the edge states in the bulk to the poles of said function
\cite{Patrick2016}.  We will assume in the following calculation that
the imaginary potential extends over a boundary region that is large
compared to the extension of the edge state such that we can assume
the potential as homogeneous, such as given for the partitioning
potential in Eq.~\eqref{eq:partitioning}.  Edge states of topological
nature should reside at zero energy.  Therefore, we start with the
ground state of the system $\ket{0}$ which fulfills $H \ket{0} = E_g
\ket{0}$ where we set the energy $E_g =0$.  On top of this, we then
can start to look for edge states with an ansatz for the wave function
$\ket{\psi} = \sum_{j=1}^N \bigl( c^\dagger_j \psi^A_j+ c_j \psi^B_j
\bigr) \ket{0}$ with $\psi^A_j,\,\psi^B_j \in \mathbb{C}$.

By applying the anti--commutation relations for the fermion ladder
operators as well as the previous relations, we can rewrite the
eigenvalue equation $H \psi = E \psi$ into two coupled recursion
relations for the coefficients $\psi^A_j$ and $\psi^B_j$:
\begin{equation}
  \Gamma_2 \psi_{j+1} + \Gamma_2^\dagger \psi_{j-1} - \Gamma_1 \psi_j = 0 \;,
\end{equation}
where $\psi_j = (\psi^A_j, \psi^B_j)^T$ and
\begin{equation}
  \Gamma_1 = 
  \begin{pmatrix} 
    \mu + \ii \gamma + E & 0 \\
    0 & - \mu - \ii \gamma + E
  \end{pmatrix}
  ,\;
  \Gamma_2 = 
  \begin{pmatrix} 
    t & -\ii\Delta \\
    \ii\Delta & -t
  \end{pmatrix} \;.
\end{equation}

Following the scheme laid out in the literature we multiply by $z^j$
where $z \in \mathbb{C}$ and sum over $j$ to rewrite the recursion
relation as
\begin{equation}
  g(z) = \left( \Gamma_2 - z \Gamma_1 + z^2 \Gamma_2^\dagger \right)^{-1} \Gamma_2 \psi_1 \;,
\end{equation}
where we defined the local generating function
\begin{equation}
  g(z) = \sum_{j} z^{j-1} \psi_j \;.
\end{equation}

Let us consider the edge of the system at $j=1$. An edge state is
expected to decay exponentially, i.e., $\psi_j \propto 1/z_1^j$ with
increasing $j$.  It can be proven \cite{Patrick2016} that poles of the
generating function $g(z)$ are positioned at the decay constants
$z_1$.  The only poles our generating function may have arise in the
determinant $\det\bigl( \Gamma_2 - z \Gamma_1 + z^2 \Gamma_2^\dagger
\bigr)$ which appears in the denominator during the calculation of the
inverse matrix.  The poles can be calculated analytically and for
zero-energy edge states $E = 0$ they are given by
\begin{equation}
  z_{a,b} = \frac{\ii \gamma + \mu + a \sqrt{ 4(- t^2  + \Delta^2) +(\mu + \ii \gamma)^2}}{2(t + b \Delta)}
\end{equation}
with $a,b \in \{ -1,1\}$.

\begin{figure}
  \centering
  \includegraphics{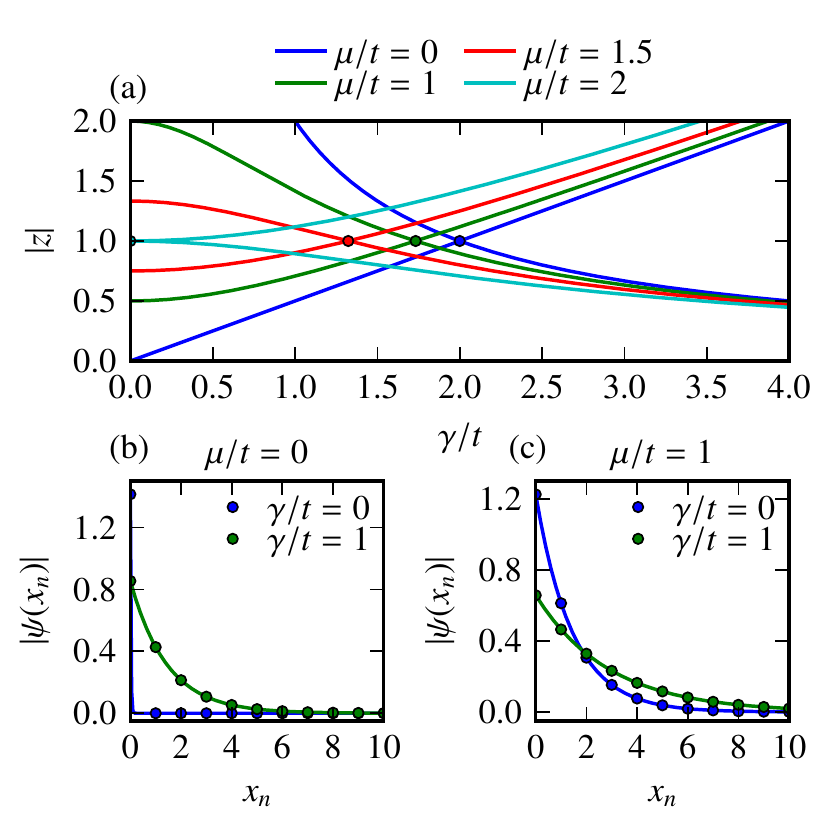}
  \caption{(a) Modulus of the poles of the generating function for $t
    = \Delta$.  The modulus of the pole gives the decay constant of
    the edge state.  Their intersection at $|z| = 1$ corresponds to
    the topological phase transition which is also found in the
    numerical calculation of the Pfaffian (filled dots).  In (b) and
    (c), the dots denote the unnormalized edge state from a numerical
    calculation with $N=100$ sites overlayed with lines for the
    analytical solution.}
  \label{fig:poles}
\end{figure}

There are four constants for each system parameter set, yet only those
greater than $1$ may contribute to a state localized at the edge
$j=1$.  Those less than $1$ correspond to the edge state that would
appear at the opposite end of the chain. But, since the complex
potential has a different sign at the other side, these solutions have
to be discarded here.  Obviously, similar results with a different
sign of $\gamma$ hold for the other edge.  A comparison to numerical
results validates the approach.  The constants of exponential decay
are given in Fig.~\ref{fig:poles}~(a).  Clearly, a large complex
potential, i.e., a large exchange of probability amplitude into or
from the system, extends the edge states into the bulk which is as it
is expected.  Observe the divergence in Fig.~\ref{fig:poles}~(b) for
$\mu=0$ and $\gamma=0$ which corresponds to perfectly localized
Majorana edge states in the Hermitian system.  At one value of
$\gamma$ at each $\mu$ the decay constants coalesce at magnitude $1$.
The notion of edge states is not valid anymore and the zero-energy
state extends into the bulk at least as far as the approximation of
constant complex potential holds true.  Interestingly, the values of
this intersection coincide with the sign change of the Pfaffian.  For
$t = \Delta $ the decay constants intersect at $1$ for
\begin{equation}
  \mu^2 + \gamma^2 = 4t^2,
\end{equation}
which is the equation that gives the phase boundary as it appeared
before in an analytical calculation of the Pfaffian invariant (see
Appendix~\ref{app:pfaffian} for a comparison with numerical results).

\section{Conculsion}

We studied the Kitaev chain, a one-dimensional topological
superconductor, hosting Majorana edge modes under the influence of
balanced gain and loss from an imaginary \PT-symmetric potential.  We
explicitly showed that the Pfaffian invariant is well defined for a
\PT-symmetric on-site potential and correctly reproduces the
topological phase behavior.  We computed the Pfaffian invariant
analytically for an alternating non-Hermitian lattice and matched the
results with numerical tight-binding calculations.

By the generating function approach we computed the decay constants of
the edge state into the bulk.  It is remarkable that despite the gain
and loss effects at the boundaries of the system, the edge state is
exponentially localized.  The decay constants depend on the gain/loss
strength $\gamma$ and the intersection of the two solutions reproduces
the phase boundary as found in numerical calculations.

The successful identification of a topological invariant in this
non-Hermitian system is a step towards a symmetry-based classification
of topological phases in non-Hermitian quantum systems.  It is also
conceivable that a similar system is realized experimentally with
optical resonators \cite{Rueter2010, Bradyn2012}.

\section*{Acknowledgement}

We thank Yuxin X.\ Zhao, Darshan G.\ Joshi, Andreas P.\ Schnyder, and
Philip M.\ R.\ Brydon for useful discussion.

\appendix

\section{Topological Invariant}
\label{app:invariant}

For the Pfaffian invariant to be useful for the class of systems
considered in this paper, it has to be well defined.  In our case of
non-Hermitian system the problem of a non-zero imaginary part could
arise such that we cannot evaluate a sign of the Pfaffian.

The extended Kitaev chain is motivated by preserving \PT~symmetry of
the non-Hermitian potential, which we will use in the following.  A
general particle-hole-symmetric matrix $H_{\text{PHS}} = - \tau_1
H_{\text{PHS}}^T \tau_1 $ can be written in Bogoliubov--de~Gennes
(BdG) form as
\begin{equation}
  H = 
  \psi^\dagger
  \begin{pmatrix}
    a & b  \\
    c & -a^T
  \end{pmatrix}
  \psi
  \quad\text{with}\quad
  b =-b^T, c = -c^T \;,
\end{equation}
where $a$, $b$, and $c$ are matrices that fulfill the given relations
and $\psi$ refers to a Nambu spinor encompassing ladder operators for
all sites.  Introducing Majorana operators will lead to new spinors
$\gamma$ and a Hamiltonian in the form
\begin{equation}
  H = \frac{\ii}{4} \gamma X \gamma \;.
\end{equation}
The matrix $X$ is skew symmetric $X^T = -X$ and takes the form
\begin{equation}
  X = 
  \begin{pmatrix}
    -i (a - a^T + b +c)& - (a + a^T - b +c)\\
    a + a^T + b -c & -i (a - a^T - b -c)
  \end{pmatrix}\;.
  \label{eq:xwithphs}
\end{equation}
If $H$ is Hermitian then $X^\dagger = - X$ and
\begin{equation}
  \Pf(X)^*  = \Pf(X^*)  = \Pf(-X^T) = \Pf(X) \;,
  \label{eq:XconjugateHermitian}
\end{equation}
such that the Pfaffian can only be a real number.

If the original Hamiltonian fulfills \PT~symmetry, each block will obey
a similar relation
\begin{equation}
  U^\dagger H^* U = H
  \implies
  u^\dagger d^* u = d \;,\qquad d\in\{a, b, c\} \;,
\end{equation}
where $u$ is one diagonal block of the unitary matrix $U$
\begin{equation}
  U =
  \begin{pmatrix}
    u & 0 \\
    0 & u
  \end{pmatrix}
  \;,
\end{equation}
which assumes that the time-reversal symmetry is chosen in such a way
that it does not mix creation and annihilation operators:
\begin{equation}
  X^* = U
  \begin{pmatrix}
    i (a - a^T + b +c)& - (a + a^T - b +c)\\
    a + a^T + b -c & i (a - a^T - b -c)
  \end{pmatrix}
  U^\dagger \;.
\end{equation}
To see how the symmetry for $H$ translates to a symmetry for $X$, we
have to apply the unitary transformation $M$ that relates the original
basis to the Majorana basis
\begin{align}
  \mathcal{H} 
  = \frac{\ii}{4} \psi^\dagger M^\dagger M  (-4\ii) H M^\dagger M \psi 
  =: \frac{\ii}{4} \gamma X \gamma \;,
\end{align}
where $\psi = (\psi_I, \psi_I^\dagger)^T$ with $\psi_I$ is the
vector consisting of all annihilation operators in the system and
$\gamma = M^\dagger \psi $ refers to the vector of Majorana operators
which has $2n$ entries.  With $I$ we denote that the indices take
values from $1$ to the number of sites $n$.  The matrix $M$ can be
written as
\begin{equation}
  \psi =
  \begin{pmatrix}
    \psi_I \\
    \psi_I^\dagger
  \end{pmatrix}
  = M^\dagger \gamma
  = \frac{1}{\sqrt{2}}
  \begin{pmatrix}
    1 & -\ii \\
    1 & \ii
  \end{pmatrix}
  \begin{pmatrix}
    \gamma_I \\
    \gamma_{I+N}
  \end{pmatrix} \;.
\end{equation}
We can rewrite the definition of $X$ into
\begin{equation}
  H = \frac{\ii}{4} M^\dagger X M \;,
\end{equation}
and use it within the relation for \PT~symmetry to get
\begin{multline}
  U^\dagger H^* U = U^\dagger \frac{-\ii}{4} M^T X^* M^* U =  \frac{\ii}{4} M^\dagger X M = H \\
  \Leftrightarrow\quad U^\dagger_M X^* U_M = - X \;,
\end{multline}
where we have used that 
\begin{align}
  M^* U M^\dagger =
  \begin{pmatrix}
    u & 0 \\
    0 & -u
  \end{pmatrix}
  = U_M \;.
\end{align}
Let us evaluate this expression for the components given in
Eq.~\eqref{eq:xwithphs} which yields
\begin{equation}
  \begin{aligned}[b]
    X^*
    &=
    \begin{pmatrix}
      -\ii (a - a^T + b +c)& - (a + a^T - b +c)\\
      a + a^T + b -c & -\ii (a - a^T - b -c)
    \end{pmatrix}^*
    \\
    &=
    U \begin{pmatrix}
      \ii (a - a^T + b +c) & -(a + a^T - b +c) \\ 
      a + a^T +b -c & \ii (a - a^T - b -c)
    \end{pmatrix} U^\dagger \;.
  \end{aligned}
  \label{eq:xstarPT}
\end{equation}

\paragraph{Treat $U$ and the signs separately}

To see what we can expect of the unitary matrix $U$ let us consider
the general skew-symmetric matrix $A$:
\begin{equation}
  \Pf(U A U^T) = \Pf(A) \det(U) = \det(u)^2 \Pf(A) = \Pf(A) \;.
\end{equation}
If the matrix $U$ is orthogonal instead and not only unitary then we
can remove it form the Pfaffian.  Should this not be the case we could
also consider that $u$ may commute with $a+ a^T$ and $b - c$ and
anticommute with $a-a^T$ and $b+c$.  After summing or substracting
these relations the following is obtained
\begin{align}
  bu + uc   & = 0, &            &  & cu +ub     & = 0, \\
  au + ua^T & =0,  & \text{and} &  & a^T u + ua & = 0.
\end{align}

If $U$ is treated by such means then we can simply consider the matrix
itself and we see that it will equal to $X$ if the diagonals vanish.
$a-a^T + b + c =0$ and $a-a^T -b -c =0$ leads to $a =a^T$, i.e., $a$
is symmetric, and $b=-c$.  If we connect this with the condition that
$u$ commutes with the blocks then we see that $u$ must commute with
$b$ and $c$ and anticommute with $a$ and $a^T$.  If the diagonals have
vanished and $U$ is removed we see from \eqref{eq:xstarPT} that $X^* =
X$ and therefore $\Pf(X)^* = \Pf(X)$.

\paragraph{Treat $U$ and the signs at the same time}

Of course there might be a way to treat $U$ and the sign differences
between $X^*$ and $X$ at the same time.  If we state the necessary
relations for each block and form differences and sums from them we
end up with
\begin{align}
  bu +uc    & = 0 & cu + ub   & = 0 \\
  ua - a^Tu & = 0 & ua^T - au & = 0 \;,
\end{align}
which will also give us a real Pfaffian.  We see that \PT~symmetry alone
is not enough.

Let us further consider the case where a non-Hermitian $H_{nH}$
potential is added to a Hermitian operator $H_{H}$, which is the case
we considered for the main part of this work.  It is also the case that
can be interpreted by particle exchange with the environment.  Our PHS
Hamiltonian is now written with a non-Hermitian on-site potential
$\tilde{a}$
\begin{equation}
  H_H + H_{nH} =
  \begin{pmatrix}
    a + \tilde{a} & b \\
    -b^* & -a^T - \tilde{a}^T
  \end{pmatrix}\;,
\end{equation}
where $b$ obeys the previously given relations.  The skew-symmetric
matrix $X_{\text{total}}$ for the whole Hamiltonian is
\begin{equation}
  \begin{aligned}[b]
    X_{\text{total}} &= X_{H} + X_{nH} \\
    &=
    \begin{pmatrix}
      -\ii (a - a^T + b -b^ *)& - (a + a^T - b -b^*)\\
      a + a^T + b +b^* & -\ii (a - a^T - b +b^*)
    \end{pmatrix} \\
    &\quad{}+
    \begin{pmatrix}
      0 & - 2 \tilde{a}\\
      2 \tilde{a} & 0
    \end{pmatrix}\;.
  \end{aligned}
\end{equation}
Now we have to apply the complex conjugation
\begin{align}
    X_{\text{total}}^* 
    = X_H + 
    U
    \begin{pmatrix}
     0 & - 2 \tilde{a}\\
     2 \tilde{a} & 0
    \end{pmatrix}
    U^\dagger
    = X_H + U X_{nH} U^\dagger,
\end{align}
where the first term originates from the Hermitian part, which makes
$X$ a real quantity, and the second part is unmodified due to two sign
changes but still subject to the matrix $U$.  The unitary matrix $U$
can be removed if it is orthogonal and commutes with $X_H$ or if it
commutes with $X_{nH}$.  The property that $U$ commutes with $X_{nH}$
is equivalent to the commutativity of $\tilde{a}$ and $u$.  But if we
recall the \PT~symmetry condition $u^\dagger \tilde{a}^* u =
\tilde{a}$ we see that this would lead to a real $\tilde{a}$.  The
assumption reduces the generally non-Hermitian part to an additional
Hermitian one.  Therefore we have to consider the first option, e.g.,
$U$ is orthogonal such that it can be removed from the Pfaffian
\begin{equation}
  \begin{aligned}[b]
    \Pf(X_{\text{total}})^*
    &= \Pf(X_H + U X_{nH} U^T) \\
    &= \Pf(X_H U U^T + U X_{nH} U^T) \\
    &= \Pf(U(X'_H  + X_{nH} )U^T) \\
    &= \Pf(X'_H  + X_{nH}),
  \end{aligned}
\end{equation}
where $X'_H = X_H$ if $[X_H ,U] = 0$ and for this case the Pfaffian is
real.  Let us go one step further in our considerations.  For the
extended Kitaev model we have a corresponding matrix $U$ that just
inverts the order of lattice sites.  The Hermitian part of this model
is isotropic in all terms except those related to pairing.  We see
that for any term of the form $c_n c_m$ an inversion of the sites
leads to $c_m c_n = - c_n c_m$ which is true for two arbitrary indices
$n \neq m$.  This sign will appear at arbitrary pairing terms,
therefore, we can remove it by a unitary transformation $W$ on $H'$
which shall have the additional minus signs compared to $H$:
\begin{align}
  W H' W^\dagger &= H \;, \\
  W &=
  \begin{pmatrix}
    \ii \mathbf{1} & 0 \\ 0 & - \ii\mathbf{1}
  \end{pmatrix} \;, \\
  W_M &:= M W M^\dagger =
  \begin{pmatrix}
    0 &  \mathbf{1} \\ -\mathbf{1} & 0
  \end{pmatrix} \;,
\end{align}
where we have introduced the orthogonal matrix $W_M$ which is $W$
expressed in the Majorana basis.  If we collect the intermediate,
steps we see that
\begin{equation}
  \begin{aligned}[b]
    \Pf(X_{\text{total}})^*
    &= \Pf(X_H + U X_{nH} U^T) \\
    &= \Pf(X_H U U^T + U X_{nH} U^T) \\
    &= \Pf[U(X'_H  + X_{nH} )U^T] \\
    &= \Pf(X'_H  + X_{nH}) \\
    &= \Pf[W_M^T (X'_H  + X_{nH})W_M] \\
    &= \Pf(X_H  + X_{nH}) \\
    &= \Pf(X_{\text{total}}),
  \end{aligned}
\end{equation}
where we have used that $W_M$ commutes with $X_{nH}$ and that
$\det(W_M)=1$.  With this it is proven that for a Hamiltonian, which
obeys PHS and has terms that are either Hermitian or \PT-symmetric
on-site potentials, the Pfaffian is real if the transformation for the
\PT~symmetry acts only as inversion on the sites.

\section{Calculation of the Pfaffian Invariant}
\label{app:pfaffian}

The invariant of the Hermitian Kitaev chain \cite{Kitaev2001} for
complex order parameter $\Delta\in\mathbb{C}$ is given in terms of the
Pfaffian.  Here, we analytically calculate the Pfaffian invariant for
an alternating potential consisting of two sublattices $A$ and $B$
experiencing gain and loss, respectively.  The Hermitian part of the
Hamiltonian is given in \eqref{eq:kitaev-real}, similar to
\cite{Kitaev2001}, by
\begin{equation}
  H_0 = \sum_n \left[ t c_n^\dagger c_{n+1} + \ii \Delta c_n c_{n+1}
    - \frac{\mu}{2} c_n^\dagger c_n + \mathrm{H.c.} \right]
\end{equation}
with the non-Hermitian potential but \PT-symmetric
potential
\begin{equation}
  U = \sum_n (-1)^n \ii \gamma c_n^\dagger c_n \;.
\end{equation}
This potential is alternating between adjacent sites and is therefore
not translationally invariant.  By dividing the system into two
sublattices $A$ and $B$ it can still be diagonalized.
\begin{widetext}
  \begin{equation}
    \begin{aligned}[b]
      H = & \sum_n \biggl[ t (c_{A,n}^\dagger c_{B,n} + c_{B,n}^\dagger c_{A,n+1})
      + \ii\Delta (c_{A,n} c_{B,n} + c_{B,n} c_{A,n+1})
      - \frac{\mu}{2} (c_{A,n}^\dagger c_{A,n} + c_{B,n}^\dagger c_{B,n})
      + \mathrm{H.c.} \biggr] \\
      & + \sum_n [ \ii\gamma c_{A,n}^\dagger c_{A,n} - \ii \gamma c_{B,n}^\dagger c_{B,n} ]
    \end{aligned}
  \end{equation}
  We now write the above Hamiltonian in the Majorana basis using the
  substitution rule
  \begin{equation}
    c_{\eta,n} = \frac{1}{2} (a_{\eta,2n-1} + \ii a_{\eta,2n}) \;,\quad
    c_{\eta,n}^\dagger = \frac{1}{2} (a_{\eta,2n-1} - \ii a_{\eta,2n})
  \end{equation}
  with $\eta = A,B$ and the Majorana operators $a$.  In this basis, the
  Hamiltonian reads as
  \begin{equation}
    \begin{aligned}[b]
      H = & \sum_n \biggl[
      \frac{\ii t}{2} ( a_{2n-1}^A a_{2n}^B - a_{2n}^A a_{2n-1}^B + a_{2n-1}^B a_{2n+2}^A - a_{2n}^B a_{2n+1}^A )
      +\frac{\ii\Delta}{2} ( a_{2n-1}^A a_{2n-1}^B - a_{2n}^A a_{2n}^B + a_{2n-1}^B a_{2n+1}^A - a_{2n}^B a_{2n+2}^A ) \\
      & -\frac{\ii\mu}{2} ( a_{2n-1}^A a_{2n}^A + a_{2n-1}^B a_{2n}^B )
      - \frac{\ii(\ii\gamma)}{2} a_{2n-1}^A a_{2n}^A + \frac{\ii(\ii\gamma)}{2} a_{2n-1}^B a_{2n}^B \biggr] \;.
    \end{aligned}
  \end{equation}
\end{widetext}
Now, we Fourier transform this Hamiltonian according to the
prescription
\begin{equation}
  a_{2n-1}^\eta = \frac{1}{\sqrt{N}} \sum_q \ee^{-\ii q n} b_{q,1}^\eta \;,\quad
  a_{2n}^\eta = \frac{1}{\sqrt{N}} \sum_q \ee^{-\ii q n} b_{q,2}^\eta
\end{equation}
where again $\eta = A,B$.  Also note that $b_q^\dagger = b_{-q}$.  In
the new basis $b_q = (b_{q,1}^A, b_{q,1}^B, b_{q,2}^A, b_{q,2}^B)^T$ we can
write the Hamiltonian in the quadratic form
\begin{equation}
  H = \frac{\ii}{4} \sum_q b_q^\dagger A(q) b_q
\end{equation}
with the skew-symmetric matrix $A(q)$.  The non-zero entries of $A(q)$
are
\begin{align}
  A_{12} &= \Delta (1-\ee^{\ii q}) \;, \\
  A_{13} &= - (\mu + \ii\gamma) \;, \\
  A_{14} &= t (1+\ee^{\ii q}) \;, \\
  A_{23} &= t (1+\ee^{-\ii q}) \;, \\
  A_{24} &= - (\mu - \ii\gamma) \;, \\
  A_{34} &= - \Delta (1-\ee^{\ii q}) \;.
\end{align}
The Pfaffian of this $4\times4$ matrix then is given by
\begin{equation}
  \Pf[A(q)] = - \Delta^2 (1-\ee^{\ii q})^2 - (\mu^2 + \gamma^2) + 2t^2(1+\cos q) \;.
\end{equation}
The topological invariant from the Pfaffian is the Majorana number
$\mathcal{M}$ which is defined in \cite{Kitaev2001} as
\begin{equation}
  \mathcal{M} = \sgn\bigl(\Pf[A(0)] \Pf[A(\pi)]\bigr)
\end{equation}
where $\mathcal{M} = -1$ indicates topologically non-trivial and
$\mathcal{M} = 1$ topologically trivial behavior.  Here, we obtain
\begin{equation}
  \mathcal{M} = \sgn\bigl[(\mu^2 + \gamma^2 - 4t^2)(\mu^2 + \gamma^2 + 4\Delta^2)] \;.
\end{equation}
The second term consists of a sum of positive numbers which is itself
always positive.  Therefore, it does not contribute to the sign of the
overall expression and the dependence on the gap parameter $\Delta$
drops out.  From the final expression for the Majorana number
\begin{equation}
  \mathcal{M} = \sgn(\mu^2 + \gamma^2 - 4t^2)
\end{equation}
we then derive the circle criterion $\mu^2 + \gamma^2 < 4t^2$ for the
topological phase presented in the main text.

\section{Partitioning Potential}
\label{app:partition}

\begin{figure}
  \centering
  \includegraphics{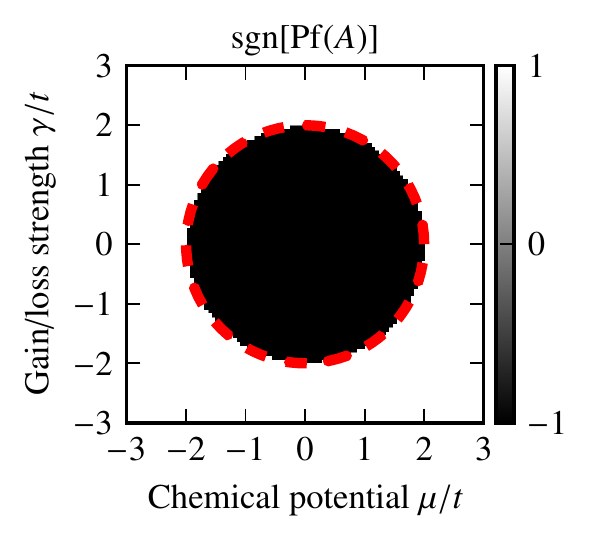}
  \caption{Topological phase diagram for the partitioning potential
    \eqref{eq:partitioning-supp} in a chain with $N=100$ sites and a
    partitioning fraction of $f=0$.  The color map indicates the sign
    of the Pfaffian, the dashed line corresponds to the analytical
    phase boundary as extracted from the generating function
    approach.}
  \label{fig:pd-partitioned}
\end{figure}

In the main text we investigate a potential which breaks Hermiticity
at both ends of the wire over the range $f$.  The potential is given
by
\begin{equation}
  \label{eq:partitioning-supp}
  U_2 = \ii \gamma \biggl[ -\sum_{n=1}^{N/2-f} c^\dagger_n c_n
  + \sum_{n=N/2+f}^{N} c^\dagger_n c_n \biggr] \;.
\end{equation}
For this potential we can numerically compute the phase diagram for a
partitioning $f=0$, i.e.\ half the wire experiences gain, the other
half loss.  This resembles the setup used for the analytical
calculation with the generating function approach.  Indeed we recover
the same criterion as derived from said method which we superimpose
with our numerical findings in figure~\ref{fig:pd-partitioned}.

\begin{figure}
  \centering
  \includegraphics{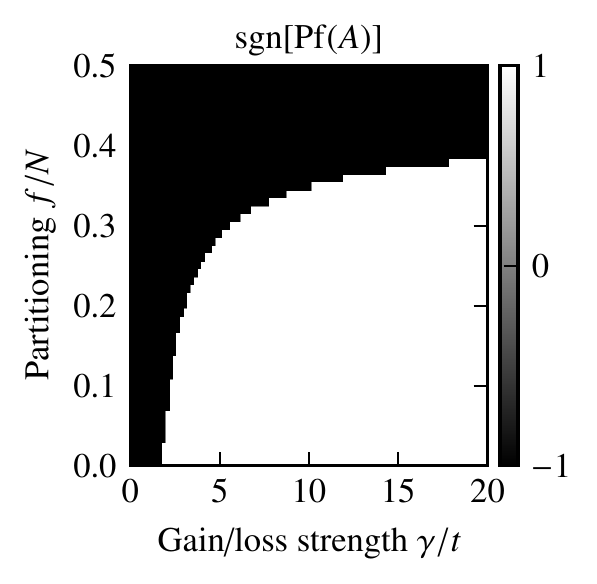}
  \caption{Topological phase diagram for the partitioning potential
    \eqref{eq:partitioning-supp} in a chain with $N=100$ sites.  A
    partitioning fraction of $f/N=0$ corresponds to half~loss\slash
    half~gain, whereas $f/N=0.5$ corresponds to the Hermitian model
    without any potential.  The colormap indicates the sign of the
    Pfaffian.  The left edge of the phase diagram corresponds to the
    Hermitian Kitaev chain.  Clearly, the phase diagram changes with
    the partitioning.}
  \label{fig:pd-partitioned-f}
\end{figure}

Because the phase diagram is again a circle one might get the
impression that the phase diagram is a circle for all \PT-symmetric
potentials.  This would require that the topological phase is
independent of the partitioning $f$.  We investigate this in a
numerical calculation of the Pfaffian and find that this is not the
case (cf.\ Fig.~\ref{fig:pd-partitioned-f}).

\begin{figure}
  \centering
  \includegraphics{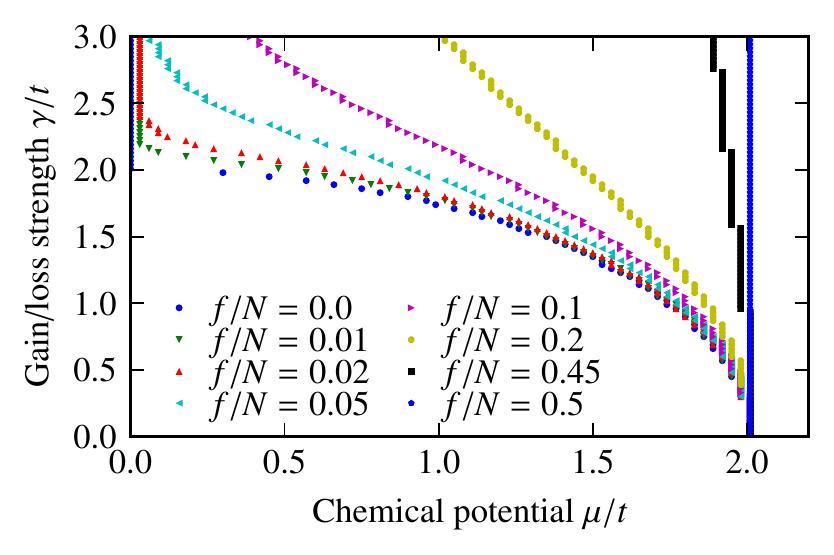}
  \caption{Numerically determined phase boundary for the potential
    \eqref{eq:partitioning-supp} at different partitioning fractions
    $f/N$ in a chain with $N=100$ sites.  With increasing partitioning
    fraction we approach the phase-boundary of the Hermitian case
    ($f/N=0.5$).}
  \label{fig:pd-boundary-f}
\end{figure}

It is interesting to study the behavior of the aforementioned
partitioning potential under change of the partitioning fraction $f/N$
to see how it changes as we approach the Hermitian limit ($f/N=0.5$).
This can be seen in Fig.~\ref{fig:pd-boundary-f}.

\section{Non-\PT-Symmetric Potentials}
\label{app:non-pt-symm}

We would like to add some comments regarding potentials which do not
fulfill \PT-symmetry based on the observations in numerical
simulations.  We assume the potential
\begin{equation}
  \label{eq:non-PT-symm}
  U = \sum_n -\ii\gamma c_n^\dagger c_n \;.
\end{equation}
This is obviously not \PT-symmetric.  However, in numerical
calculations we nevertheless find edge modes and were able to extract
a rule of thumb for the topological invariant.  As in the
\PT-symmetric case the edge modes are characterized by a vanishing
real and imaginary part of the eigenvalues.  In the topological regime
the Pfaffian is real and has a sign of $-1$, correctly indicating
topological behavior.  In the trivial regime, however, the Pfaffian
becomes imaginary and the sign of the real part is no longer strictly
positive, making the sign of the real part alternate.  Therefore in
this case we extended the condition for a topological phase to the
following statement.  For a topological phase to exist, the Pfaffian
must be real and its sign must be negative.  Imaginary Pfaffian or a
positive sign of the Pfaffian corresponds to the trivial phase.

\begin{figure}
  \centering
  \includegraphics{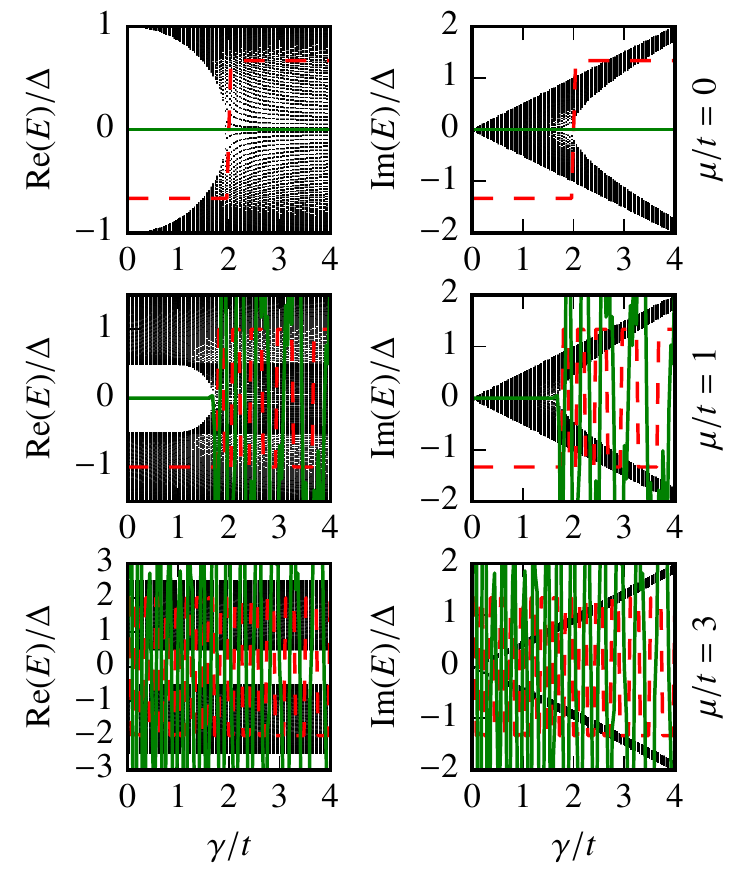}
  \caption{We show spectra for the non-\PT-symmetric
    potential in Eq.~\eqref{eq:non-PT-symm} in a chain with
    $N=100$ sites at different chemical potential over the gain/loss
    strength $\gamma$.  The dashed line corresponds to the sign of the
    real part of the Pfaffian, the solid line to the value of the
    imaginary part.}
  \label{fig:pure-loss}
\end{figure}

For the potential in Eq.~\eqref{eq:non-PT-symm}, we show the
eigenvalue spectra overlayed with the sign of the real part and the
bare value of the imaginary part of the Pfaffian in
Fig.~\ref{fig:pure-loss}.

\begin{figure}
  \centering
  \includegraphics{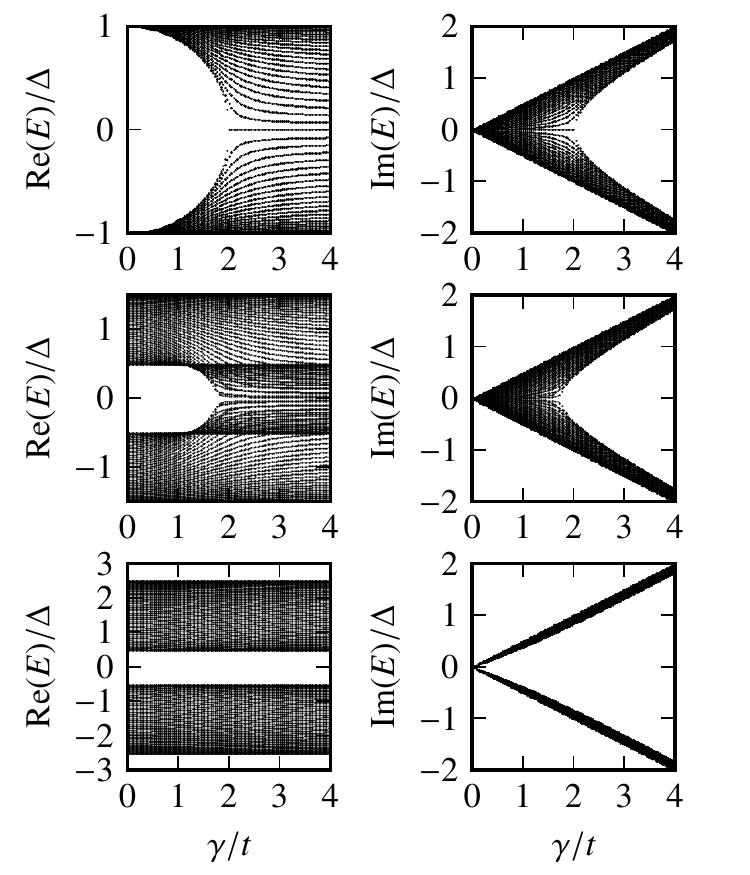}
  \caption{We show bulk spectra for the non-\PT-symmetric
    potential in Eq.~\eqref{eq:non-PT-symm} in a chain with
    $N=100$ sites at different chemical potential over the gain/loss
    strength $\gamma$.}
  \label{fig:pure-loss-pbc}
\end{figure}

As described in the main text for the potential $U_1$, the bulk
spectrum for this system differs qualitatively from its open
counterpart (see Fig.~\ref{fig:pure-loss-pbc}).  In the case
$\mu/t=0$ the real part of the spectrum becomes gapless in the
topologically trivial phase.

This does also not conform with analytical predictions.  If we
calculate the Pfaffian of the present model we find
\begin{equation}
  \Pf[A(q)] = 2 t \cos q - (\mu + \ii \gamma) + 2 \ii \Delta \sin q \;.
\end{equation}
Simply evaluating the condition for the Majorana number would in this
case yield
\begin{equation}
  \begin{aligned}
    \mathcal{M} &= \sgn\bigl(\Pf[A(0)] \Pf[A(\pi)] \bigr) \\
    &= \sgn\bigl( \mu^2 - 4 t^2 - \gamma^2 + 2 \ii \gamma \mu \bigr) \;.
  \end{aligned}
\end{equation}
The argument of the sign function is complex and the result is thus
undefined.  Even if we apply the same demands as for the real-space
tight-binding results, namely that the imaginary part of the Pfaffian
vanishes, this would lead to a contradiction with the numerical
results.  Here this would mean that $\gamma\mu=0$ which implies that
either $\gamma=0$ or $\mu=0$.  In Fig.~\ref{fig:pure-loss}, however,
we find that for $\mu/t=1$ the topological phase transition does not
take place at $\gamma=0$ as would be required by the ill-defined
argument.  We hence conclude that the Pfaffian is not a good
topological invariant for non-\PT-symmetric systems.

\end{document}